\documentclass[%
 rsi,
 amsmath,amssymb,
reprint,%
]{revtex4-2}
\pdfoutput=1
\usepackage{graphicx}% Include figure files
\usepackage{dcolumn}% Align table columns on decimal point
\usepackage{bm}% bold math
\usepackage {subfigure}

\usepackage[utf8]{inputenc}
\usepackage[T1]{fontenc}
\usepackage{mathptmx}
\usepackage{etoolbox}
\usepackage{hyperref}

\makeatletter
\def\@email#1#2{%
 \endgroup
 \patchcmd{\titleblock@produce}
  {\frontmatter@RRAPformat}
  {\frontmatter@RRAPformat{\produce@RRAP{*#1\href{mailto:#2}{#2}}}\frontmatter@RRAPformat}
  {}{}
}%
\makeatother
\begin{document}

\preprint{AIP/123-QED}

\title[]{Amplification of arbitrary frequency chirps of pulsed light on nanosecond timescales}
% Force line breaks with \\

\author{B.S. Clarke*} 
\author{P.L. Gould}
\affiliation{ 
Department of Physics, University of Connecticut, Storrs, Connecticut 06269, United States%\\This line break forced with \textbackslash\textbackslash
}%

\date{\today}% It is always \today, today,

\begin{abstract}
We have developed a system for producing amplified pulses of frequency-chirped light at 780 nm on nanosecond timescales. The system starts with tunable cw laser light and employs a pair of fiber-based phase modulators, a semiconductor optical amplifier, and a tapered amplifier to achieve chirp rates exceeding 3 GHz/10 ns and peak powers greater than 1 W. Driving the modulators with an arbitrary waveform generator enables arbitrary chirp shapes, such as two-frequency linear chirps. We overcome the optical power limitations of the modulators by duty cycling and avoid unseeded operation of the tapered amplifier by multiplexing the chirped pulses with “dummy” light from a separate diode laser.
\end{abstract}

\email[Author to whom correspondence should be addressed: ]{bradley.clarke@uconn.edu}

\maketitle

\section{Introduction}

Since lasers are ubiquitous in modern science and technology, manipulation of their output light fields is critically important \cite{Nag2014}. Among the properties which can be controlled are amplitude (or intensity), and phase (or frequency). The time scales for control cover a wide range, depending on the application. High-speed modulation is generally done in the frequency domain, due to electronic speed limitations. On the other hand, modulation at lower speeds would require very high frequency resolution, and is thus better suited to control in the time domain. For frequency-domain control with a pulse shaper, short optical pulses have their broad bandwidth spatially dispersed, enabling the amplitudes and phases of the various frequency components to be individually manipulated. These components are then reassembled, yielding a modified pulse \cite{Weiner2000,Monmayrant2004}. In the time-domain, control can be accomplished using acousto-optical modulators (AOMs) \cite{Thom:13} and electro-optical modulators (EOMs) driven by RF waveforms. The former can provide pulses of light with durations down to 10s of ns and frequency shifts typically in the 100 MHz range. The latter can modulate either amplitude (intensity modulators) or frequency (phase modulators). Fiber-based EOMs utilize a waveguide of electro-optical material (e.g., lithium niobate) and can operate at speeds up to 10s of GHz with relatively low drive voltages. Modulation in the 10-100 ps range is difficult to achieve with either frequency-domain or time-domain techniques.

There are several ways one can produce frequency-chirped light in the time domain. One method is to vary a frequency-controlling parameter of the laser itself in real time. Direct modulation of the injection current of an external-cavity diode laser (ECDL) is one such example, though finite system response limits the modulation speed. In addition, ramping the injection current leads to undesirable intensity modulation, which can be minimized by using the frequency-chirped light to injection lock a separate free-running diode laser (FRDL) \cite{Wright2004}. Instead of modulating the laser itself, one can produce the frequency variation by modulating the light externally, after it has left the laser \cite{Rogers2007,Teng2015,Rogers2016,Kanno2010,Wang2015}. This is the path we have followed here, using fiber-based electro-optical phase modulators (PMs) driven by an arbitrary waveform generator (AWG) to produce frequency-chirped light at 780 nm. These chirped pulses are then double-stage amplified by a semiconductor optical amplifier (SOA) followed by a tapered amplifier (TA).  Previous work along these lines \cite{Rogers2007} utilized a PM in a multipass fiber loop, followed by injection locking of a FRDL, to produce pulses of arbitrary frequency-chirped light. The pulse repetition rate was limited by the fiber loop time, and the output power was limited to that of the FRDL. Another related scheme \cite{Rogers2016} used a combination of fiber-based phase and intensity modulators, followed by a double-pass tapered amplifier with a fiber delay line. The system described here has several advantages: it avoids risks of optical damage associated with a double-pass TA; the repetition rate is not limited by fiber delays; and the peak output power of the chirped pulses can exceed 1 W.

There are many applications of chirped pulses for which the system described here should prove useful. In atomic and molecular systems, chirped light can provide efficient population transfer \cite{Vasilev2005,Torosov2021,Dou2016}. In the adiabatic regime, the transfer efficiency can approach 100\% and can be robust with respect to pulse parameters. Competition with spontaneous emission from excited states, which typically occurs on the 10s of ns timescale, requires pulses and chirps of short duration. Specific applications employing chirped-pulse population transfer include: driving Raman transitions \cite{Djotyan2004,Collins2012,Liu2014}; manipulation of atomic spin-waves \cite{He2020,He2020a}; laser cooling of atoms via sawtooth wave adiabatic passage \cite{Norcia2018,Greve2019}; strong optical forces on atoms \cite{Miao2007,Metcalf2017} and molecules \cite{Jayich2014}; light-induced ultracold collisions \cite{Wright2005,Wright2007,Pechkis2011,Wright2015}; ultracold molecule formation by photoassociation \cite{Carini2013,Carini2015}; and high-speed spectroscopic gas sensing \cite{Lou2019}.
\vspace{-3mm}

\section{Apparatus}
\subsection{Optical Setup}
\vspace{-3mm}

\begin{figure}
\includegraphics[width=90mm,scale=0.75]{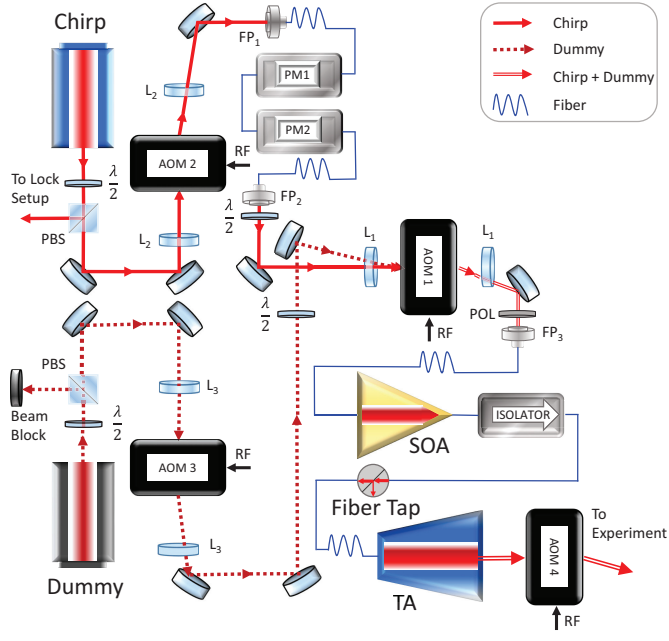}
\caption{\label{fig:opt}Chirp system optics diagram. Phase modulated Chirp light is multiplexed with Dummy light, where both are sent through AOM1 and are jointly coupled to fiber at port FP3, followed by SOA and TA amplification. Additional information pertaining to the roles of the Chirp, Dummy, AOMs, PMs, SOA, and TA are described in the text. Symbols: $\lambda/2$, half-wave plate; PBS, polarizing beam splitter; L, lens; FP, fiber port; POL, polarizer. }
\label{fig:optics}
\end{figure}

The low optical power handling capability of present-day lithium niobate phase modulators at 780 nm is an unfortunate technical limitation toward realizing complex frequency-chirp shapes at moderate to high optical powers; however, we have developed a system to circumvent this problem. As shown in Fig. 1, our chirp system is comprised of two lasers and two laser amplifiers: a Toptica DL Pro ECDL (Chirp), a Hitachi HL7852G FRDL (Dummy), an Innolume SOA-780-20-YY-30dB (SOA), and a Toptica BoosTA Pro (TA). The Chirp laser provides the light to be frequency-chirped and amplified, where a small portion of its output is used for frequency locking to an optical cavity. The rest of its output is directed to a pair of fiber-coupled PMs in series. To avoid PM optical damage, the time-averaged optical power sent through these devices is amplitude modulated by AOM2 (80 MHz) as seen in Fig. 1. The phase modulated (frequency-chirped) light to be amplified is then sent to AOM1 (200 MHz), whose -1st order output is coupled to fiber port FP3. Pulsed -1st order AOM1 output is then amplified by the fiber-coupled SOA which, after passing through a fiber isolator and fiber-tap power monitor, seeds the fiber-coupled TA. 

We note that preamplification by the SOA is necessary to take full advantage of the high power capability of the TA. This is due to the time-averaged power limitation (5 mW, or 7 dBm) of the PMs in addition to high insertion losses, 2.5 dB and 3.3 dB for PM1 and PM2, respectively. When accounting for additional losses associated to free-space to fiber and fiber-to-fiber coupling, the overall loss between FP1 and FP2 is $\sim$11 dB. Ultimately, the peak FP2 output of the Chirp-PM chain (6 dBm) is far below the moderate seed power ($>$ 13 dBm) required by the TA to realize a 1 W (30 dBm) output. This necessitates the use of an  intermediate optical amplification stage, such as an SOA. Since our SOA has a 780 nm low-signal gain of roughly $+$29 dB with an injection current of 350 mA, FP2's free-space output is reduced to the appropriate SOA seed level by controlling the RF amplitude sent to AOM1 in addition to the combined use of a $\lambda /2$ waveplate and a polarizer, as depicted in Fig. 1.

\begin{figure}
\includegraphics[width=90mm,scale=0.75]{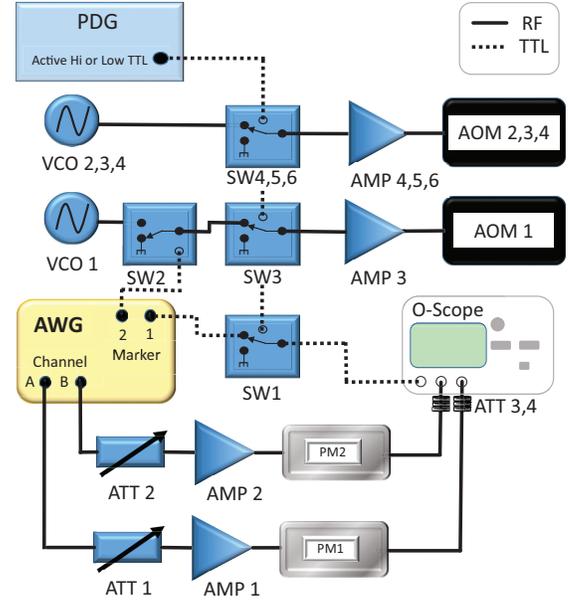}
\caption{\label{fig:elec}Chirp system electronics diagram. A computer (not shown) is used to program a dual channel arbitrary waveform generator (AWG) with user defined waveforms for the PMs. All five pulse delay generator (PDG) signals are synchronized, with the various switches (SW) being active when the PDG TTL signals are either high (SW1,3,4,6) or low (SW5). See text for additional details.}
\end{figure}

To prevent TA thermal damage, it is important to maintain seeding of the TA when the chirped light is not present. To this end, we multiplex Dummy with Chirp light using AOM3 (80MHz). The two beams, which are alternated in time, are aligned such that AOM1 0th order Dummy light is perfectly overlapped with AOM1 -1st order Chirp light, after which both beams are then fiber coupled at fiber port FP3. The important consequence of this alignment scheme is that the light in the fiber is exclusively Dummy light when the AOM1 drive is off, and primarily Chirp light when the AOM1 drive is on. After amplification by the SOA and TA, and subsequent beam shaping with free-space optics, the Dummy/Chirp light is incident on AOM4 (40MHz). The light incident on AOM4 is unfocused in order to prevent optical damage by the high-power TA output. By synchronizing the drive for AOM4 with that for AOM1, we can  eliminate the majority of the Dummy light, sending only the chirped light downstream to the experiment. 

In many applications, the presence of far-detuned Dummy light and/or amplified spontaneous emission (ASE) is harmless; however, residual Dummy/ASE light can be problematic in some cases. For example, in experiments \cite{Carini2015} producing ultracold Rb$_{2}$ molecules by photoassociation of ultracold atoms, the driving of various molecular transitions can lead to photodissociation. If the Dummy’s wavelength is tuned sufficiently far (e.g., $\sim$8 nm in our case) from the Chirp’s, such complications due to residual Dummy light and/or broadband ASE can be circumvented through the use of a narrow-band optical filter.  

In order to accurately characterize the light sent to the experiment, a small portion of the TA's filtered output is used for diagnostic purposes. The TA amplified chirped pulses are combined with reference light from a separate ECDL on a 9.5 GHz amplified photodiode (Thorlabs PDA8GS) and the resulting heterodyne signal (Fig. 3(b.i)) is sent to an 8 GHz oscilloscope for averaging. The amplitude profile of the pulse (Fig. 3(b.ii)) and its time-dependent frequency (Fig. 3(b.iv)) are determined using peak-to-peak analysis. The spectrograms in Fig. 4(a-c) are obtained using a short-time Fourier transform (STFT) algorithm \cite{NISpectrogram}.

\subsection{Electronics}

The chirp system’s electronics setup is shown in Fig. 2. A computer (not shown) is used to program an 8 GS/s Euvis AWG 872 with user-defined RF waveforms for channels A and B. USB digital step attenuators (Telemakus TEA4000-7) ATT1 and ATT2 are used to precisely set the amplitudes of the waveforms before amplification. High-gain 6 GHz bandwidth precision RF amplifiers (Hittite HMC-C075), AMP1 and AMP2, are used to amplify the waveforms for driving our EOSpace PMs (PM-5K3/5SES-10-PFA-PFA-780-LV-UL), PM1 and PM2 respectively. By attenuating (with ATT3 and ATT4) and 50-ohm terminating the amplified RF signals for our externally terminated PMs, we are able to monitor the programmed arbitrary waveform RF output on an 8 GHz oscilloscope for diagnostic purposes. The four AOMs in Figures 1 and 2 are driven by voltage-controlled oscillators (VCO) and amplifiers (AMP). The RF signals are modulated by RF switches (SW) whose timings are controlled by a BNC 577 pulse delay generator (PDG) or a marker pulse from the AWG. Two AND gates are used to synchronize the timing of RF and optical signals. The first AND gate is formed by SW1, which is fed TTL signals from the PDG and AWG, and is used to trigger two 8 GHz oscilloscopes for RF diagnostics and optical heterodyne measurement. The second AND gate, which ultimately controls SOA and TA seeding through AOM1, is formed by SW3 switching the output of AWG triggered SW2. The AWG Marker 2 signal sent to SW2 is synchronized with the RF signals that are sent to the PMs. 

The multiplexing of Chirp and Dummy light is controlled by AOM2 and AOM3 and switches SW4 and SW5, where the Chirp light is sent through PM1 and PM2 toward AOM1, and Dummy light is blocked or set at a reduced level. This multiplexing of Chirp and Dummy light is crucial for safe and reliable operation of the TA, where extended absence of seed light could result in irreparable damage to the device. Note, however, that TAs can be operated un/under-seeded for brief periods of time without harm, for example, during the $\sim$100 ns intervals between chirped pulses.

\begin{figure*}
  \centering
  \begin{subfigure}[]
	\centering
	\includegraphics[width=87mm,scale=0.5]{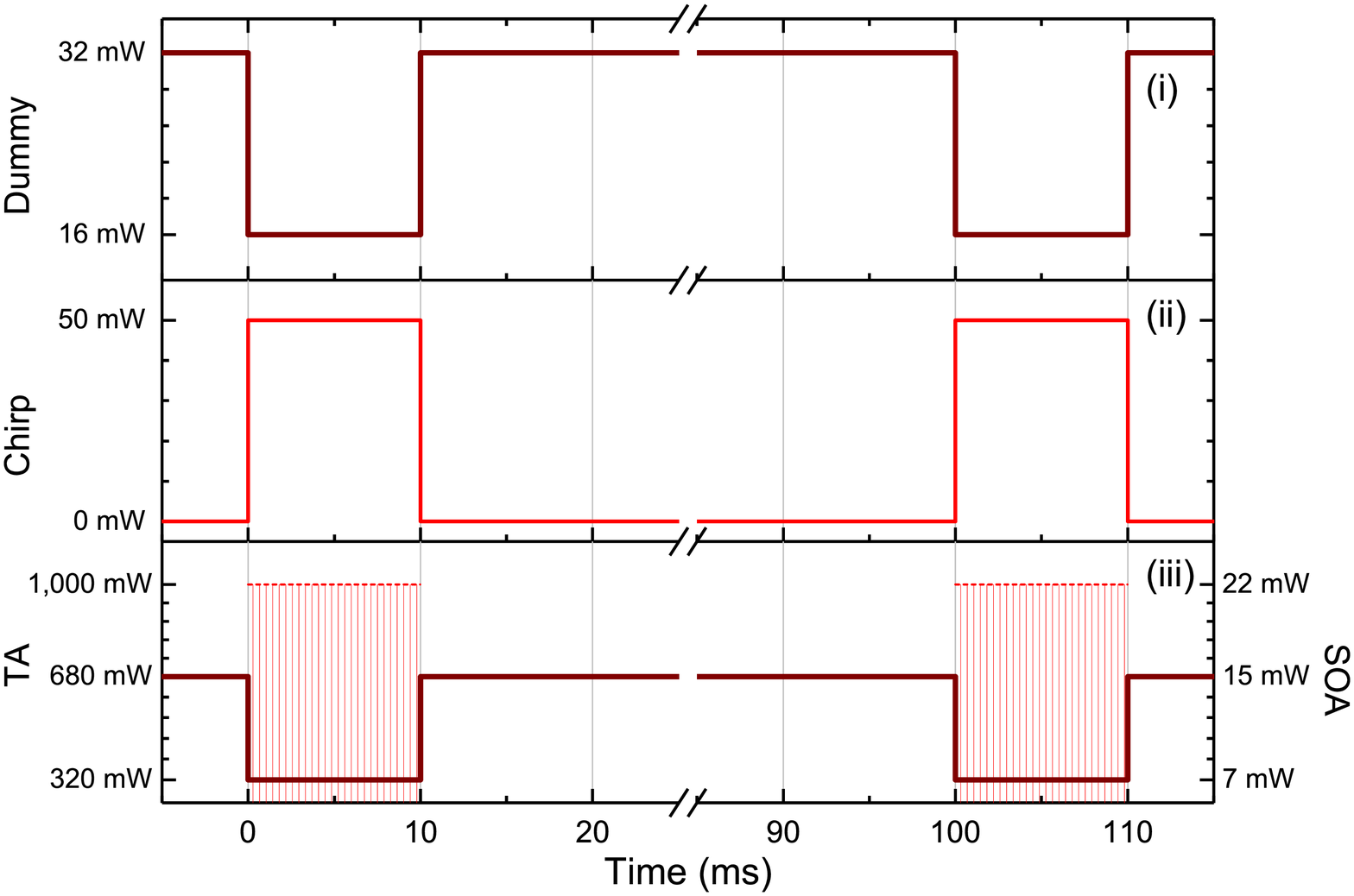}
	\label{fig:pulses}
  \end{subfigure}
  \hfill
  \begin{subfigure}[]
	\centering
	\includegraphics[width=87mm,scale=0.5]{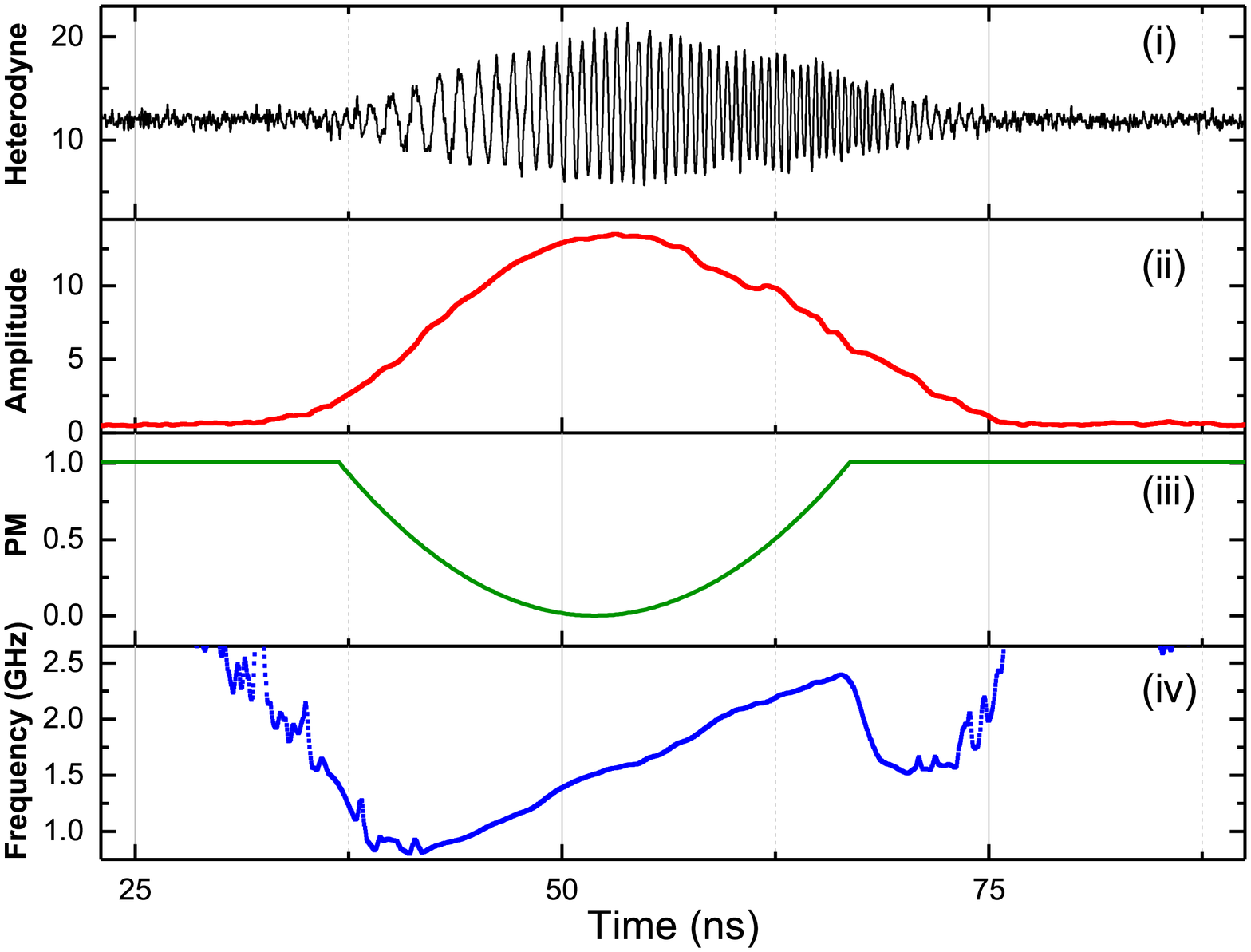}
	\label{fig:linearall}`
  \end{subfigure}
  \caption{\label{fig:fig1} (a) Chirp system timing diagram with a 100 ms repetition period where an example chirp window of 10 ms is shown. At $t=0$, Dummy light is set to a reduced level by AOM3 (a.i) while Chirp light exiting AOM2 is sent to the PMs (a.ii). The pulses (red vertical lines) within the chirp window (between 0-10 ms and 100-110 ms) in (a.iii) represent AOM1-modulated Chirp light sent to the SOA and TA. Note that only a small fraction of the many short pulses (100,000) is shown within the chirp windows depicted in (a.iii). The solid dark curve in this plot represents amplified Dummy light. The vertical axes for the Chirp, Dummy, TA, and SOA plots represent the peak powers after AOM2, AOM3, AOM4, and at the fiber tap respectively. (b) Characteristics of a positive linear frequency chirp. The amplitude (b.ii) and instantaneous frequency (b.iv) of the chirp are averages from 50 successive heterodyne (b.i) measurements, where the Chirp laser is locked 1.5 GHz blue of the reference. The waveform depicted in (b.iii), described mathematically by Eq 2, represents the quadratic waveform programmed into both AWG channels for PM1 and PM2. Except for the frequency plot (b.iv), vertical axis units are arbitrary.}
\end{figure*}

\subsection{Pulse Timing}

A typical pulse sequence for our setup is shown in Fig. 3(a). The repetition period for the sequence is 100 ms, where the Dummy seeds the SOA/TA for no less than 90 ms. The maximum chirp duration is restricted to 10 ms (Fig. 3(a.ii)), ensuring that the time-averaged Chirp power sent to PM1 stays below its 5 mW optical damage threshold. With Dummy light set to a reduced level by AOM3 at t = 0 ms in Fig. 3(a.i), -1st order AOM1 Chirp light is modulated to produce pulses of frequency-chirped light (Fig. 3(a.iii)), where the pulse width is set by AWG Marker 2 and is limited by the rise/fall times of SW2 (MiniCircuits ZASWA-2-50DR+) and AOM1 (Brimrose TEM-200-50-780), which are 5 ns and 10 ns (for a 50 µm beam diameter) respectively. As seen in Fig. 3(a.iii), the SOA/TA combination is always seeded by either Dummy light or chirped pulses.

The properties of a typical chirped pulse are shown in Figs. 3(b.i-iv). This 30 ns duration positive linear chirp with a repetition period of 100 ns is repeated 100,000 times during the 10 ms chirp window of Fig. 3(a.iii). As discussed in Sect. II.D, the time-varying frequency is the time derivative of the phase, so a linear chirp results from a quadratic voltage applied to the PMs, which is shown in Fig. 3(b.iii). The AWG programming includes a temporal offset between channels A and B to compensate for electrical and optical delays associated with series-connected PM1 and PM2. One single heterodyne measurement arising from the superposition of the chirped pulse and the cw reference laser is shown in Fig. 3(b.i). The amplitude vs. time (Fig. 3(b.ii)) and frequency vs. time (Fig. 3(b.iv)) plots are extracted and averaged from 50 heterodyne measurements. We note that apparent frequency fluctuations in Fig.3(b.iv), apart from the low amplitude regions, are likely due to residual amplitude modulation (RAM) \cite{Diehl2017} from imperfect polarization matching in the polarization-maintaining fibers, where the effect is exacerbated by high RF-PM drive power. 

\subsection{Phase Modulation}

Since the total phase angle of a light field with carrier frequency $f_0$ and time-dependent PM-induced phase shift $\phi(t)$ is given by $\Phi(t)=2\pi f_0+\phi(t)$, and since the instantaneous frequency of a light field is equal to the time derivative of the total phase angle, the time-dependent frequency of phase modulated light is given by

\begin{eqnarray}
f(t)=\frac{1}{2\pi}\frac{d\Phi(t)}{dt}=f_0+\frac{1}{2\pi}\frac{d\phi(t)}{dt}
\label{eq:freq}
\end{eqnarray}

\noindent The maximum phase shift that can be realized for any given PM is ultimately determined by two parameters: the voltage required by the PM to produce a $\pi$ phase shift $\left(V_{\pi}\right)$ and the maximum RF power handling of the PM-RF amplifier chain. To help elucidate the following discussion, consider the voltage waveform $V(t)$ required to produce a positive linear frequency chirp, such as the one shown in Fig. 3(b.iv):

\begin{eqnarray}
\phi(t)=\frac{\pi}{V_{\pi}}V(t)=\frac{\pi}{V_{\pi}}\frac{4 V_p}{\tau^2}\left(t-t_0\right)^2
\label{eq:volt}
\end{eqnarray}

\noindent where the waveform is centered about $t_0$, has duration $\tau$, and cycles between 0 and peak voltage $V_p$ twice. Since a chirp's frequency range ($\Delta f$) is dependent on phase modulation duration ($\tau=\Delta t$), the capabilities of a PM-RF system are best characterized through the time-bandwidth product, $\Delta f \Delta t$. For maximum realizable peak voltage $V_{p \hspace{0.5mm} max}$, the maximum time-bandwidth product for a linear frequency chirp is

\begin{eqnarray}
\left(\Delta f \Delta t\right)_{max}=4 \frac{V_{p \hspace{0.5mm} max}}{V_\pi}
\label{eq:tbp}
\end{eqnarray}

With a maximum AWG peak output voltage of 635 mV, a sub-1 GHz Hittite amplifier gain of +26 dB, 1.5 dB attenuation between the AWG and RF amplifier, and $V_{\pi}=1.8 \hspace{0.5mm}V$, the time-bandwidth product for a linear chirp produced by one of our PMs is $23.7$. This value is consistent with the measured frequency range of $\sim 1.6$ GHz for the linear chirp shown in Fig. 3(b), where a quadratic waveform 30 ns in duration with 1.5 dB attenuation was applied to both PM-RF amplifier chains. Due to the finite frequency response of the chirp system's RF components, a linear chirp's time-bandwidth product is reduced at high chirp rates. Nevertheless, with 1 dB of RF attenuation, we have realized a linear chirp of 3 GHz in 10 ns, corresponding to a time-bandwidth product of 15 for each PM.

\section{Results}
\subsection{Arbitrary frequency chirps}

By appropriate programming of the AWG, a variety of chirp shapes can be realized. Some examples are shown in Fig. 4. These are displayed as spectrograms \cite{NISpectrogram}, which are contour plots of short-time Fourier transforms (STFTs) of the heterodyne signals. The color coding denotes the relative intensity of the various frequency components at the indicated time. For all three examples, the intensity pulse has a FWHM of 22 ns. Fig. 4(a) is a positive linear chirp spanning 1.6 GHz in 30 ns, similar to that shown in Fig. 3(b). The sum of a slow linear chirp and an arctan chirp is shown in Fig. 4(b). This shape would be particularly useful for driving a Raman transition, where a slow chirp sweeps adiabatically through the first leg of the transition, followed by a rapid frequency jump to minimize the time spent in the excited state, concluding with another slow chirp to adiabatically drive the final transition. Fig. 4(c) displays a two-frequency chirp in which two offset frequencies are chirped in parallel. To realize this, a 1.0 GHz sinusoid is applied to PM1 with the modulation depth, determined by ATT1, chosen to suppress the carrier. This yields the +1 and -1 order sidebands, separated by 2.0 GHz. A quadratic voltage is applied to PM2, which provides a linear chirp to each of the sidebands. Similar to the linear-plus-arctan case, the two-frequency chirp would prove useful for Raman transitions, but without requiring the rapid jump, and with the added benefit that the timing of passing through the two resonances can be adjusted by changing the offset between the two chirped frequencies. This example highlights the flexibility offered by using two PMs. The same voltage can be applied to both PMs, as done for Figs. 4(a) and 4(b), or each PM can be driven independently, as for the two-frequency chirp in Fig. 4(c).

\begin{figure}
\centering
\includegraphics[width=90mm]{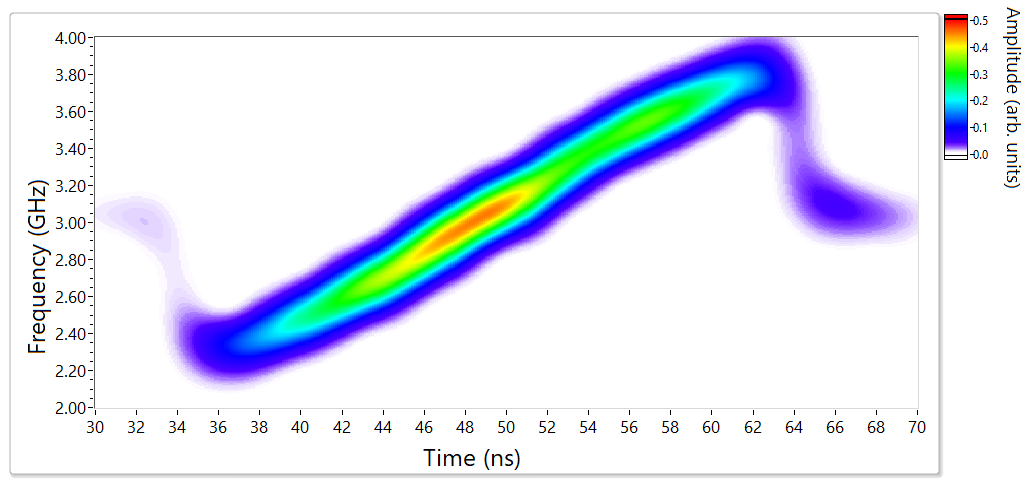}
\includegraphics[width=90mm]{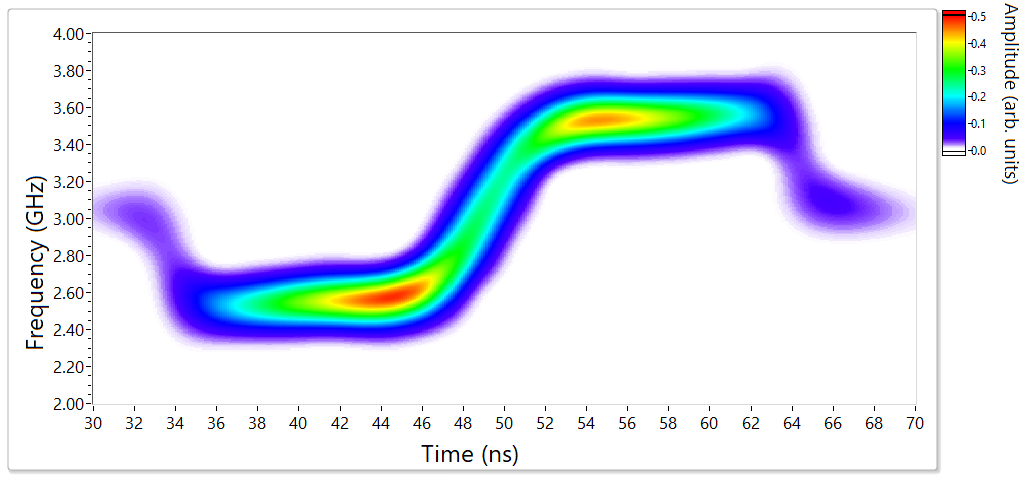}
\includegraphics[width=90mm]{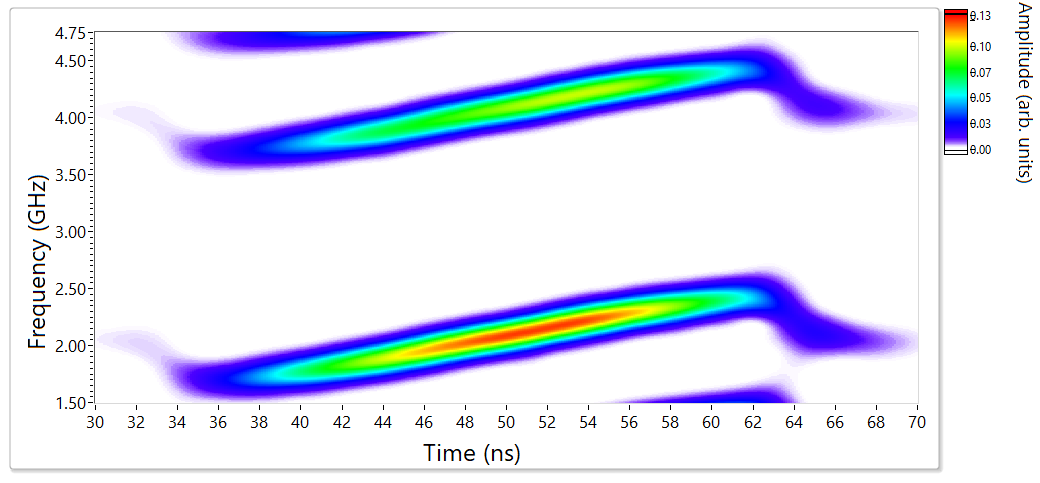}

\caption{\label{fig:atp}Short-time Fourier transform (STFT) spectrograms of three arbitrary frequency-chirps. The carrier frequency for all chirps is 3.0 GHz. (a) Linear, (b) ArcTan+Linear, and (c) Two-Color Linear.}
\end{figure}

\subsection{Multispectral power considerations}

When driving transitions with chirped light, the faster the chirp, the greater is the intensity required to maintain adiabaticity. Therefore, higher peak powers will allow rapid and robust population transfer. The output power from the TA is determined by its injection current, I, and the power with which it is seeded. We typically operate with I = 3.0 A along with Chirp and Dummy seeded SOA powers of 22 mW and 15 mW, respectively, as shown in Fig. 3(a). Under these conditions, the peak output power of Chirp and Dummy light during the chirp window are 1 W and 320 mW, respectively. Since the Chirp pulses are superimposed on amplified Dummy light, a narrowband filter can be used to block Dummy light if its wavelength is tuned sufficiently far from the Chirp's. With the Dummy tuned 8 nm red of the Chirp's wavelength, a filter centered at 780 nm with a 3 nm FWHM (AVR Optics LL01-780-12.5) provides $>$60 dB attenuation of TA amplified Dummy light. This filter also eliminates a large fraction of the broadband ASE, which would reduce the driving of unwanted transitions, especially in applications to molecules.

Since our chirp setup employs double-stage optical amplification, the contribution of ASE to the TA's output must be considered. To determine how much of the total output power of the TA can be attributed to ASE, we use a heated Rb vapor cell to absorb the resonant Chirp light, while allowing passage of the broadband ($\sim$12 nm FWHM) ASE. With a Rb vapor cell, 5 cm in length, heated to 65 $^{\circ}$C, low-intensity light within 400 MHz of the $^{85}$Rb (F=3 -> F’=4) transition will be >99\% absorbed \cite{McCarron2007}. Sending attenuated (sub-saturation intensity) TA output through the cell, with resonance-locked Chirp light solely seeding the SOA and TA, we measure 3\% transmission, meaning that $<$3\% of the TA's output can be attributed to ASE.

\section{Conclusion}
There are many ways in which the technique described here could be extended and improved. Different wavelengths could be employed, contingent on the availability and performance of fiber-based EOMs, SOAs, and TAs or fiber amplifiers. EOMs capable of handling higher optical powers, such as those using lithium niobate with MgO doping, would allow a simpler setup. Higher peak output powers, while sacrificing repetition rate, could be obtained using additional stages of amplification, e.g., with a multi-pass pulsed Ti-sapphire amplifier \cite{Seiler2005}. Shorter pulses could be realized by incorporating a fiber-based electro-optical intensity modulator in series with the phase modulators \cite{Rogers2016}. Higher speed electronics would enable faster chirp rates and larger chirp ranges. Extended chirp ranges could also be achieved by using the serrodyne technique \cite{Rogers2011} to suddenly shift the frequency during the chirp. Recent developments in optical spectral engineering \cite{Holland} could potentially yield more faithful production of desired chirps by compensating for imperfections in electrical and optical system responses.

In conclusion, we have demonstrated a versatile system for generating high-power pulses of frequency-chirped light at 780 nm. The use of two independent phase modulators enables flexibility in the chirps which can be created. Various chirp shapes have been demonstrated, such as linear, linear plus arctan, and two-frequency linear. Two stages of amplification, in conjunction with multiplexed seeding with Dummy light, result in high output power without exceeding the damage limits of the modulators or the tapered amplifier. These high-power chirped pulses are well suited to rapidly and robustly drive transitions in atoms and molecules, and should therefore find use in areas such as atom optics, ultracold molecule formation, Raman transitions, and spectroscopic sensing. 

\begin{acknowledgments}
This work was partially supported by the National Science Foundation, Grant No. PHY-1506244. We thank D.J. McCarron for useful discussions.
\end{acknowledgments}

\vspace{10mm}

\section*{author declarations}
Authors have no conflicts to disclose.

\section*{data availability}

The data that support the findings of this study are available from the corresponding author upon reasonable request.

\nocite{*}
\bibliography{chirpv7}% Produces the bibliography via BibTeX.

\end{document}